\documentclass[sn-mathphys]{sn-jnl}
\usepackage{float}
\usepackage{lineno,hyperref}
\usepackage{amsmath}
\usepackage{subfigure}
\usepackage{amssymb}
\usepackage{dsfont}
\usepackage{color}
\usepackage{graphicx}
\usepackage{array}
\usepackage[utf8]{inputenc}
\modulolinenumbers[5]
\begin{document}
\title[Vortex-lattice formation in SO-coupled spin-2 BEC under rotation]{Vortex Lattice Formation in Spin–Orbit‑Coupled Spin‑2
Bose–Einstein Condensate Under Rotation}
\author*[]{Paramjeet Banger} \email{2018phz0003@iitrpr.ac.in}
\affil[]{Department of Physics, Indian Institute of Technology Ropar, Rupnagar 140001, Punjab, India}

\abstract{We investigate the vortex-lattice configuration in a rotating spin orbit-coupled spin-2 Bose-Einstein condensate confined in a quasi-two-
dimensional harmonic trap. By considering the interplay between rotation 
frequency, spin-orbit couplings, and interatomic interactions, we explore a variety of 
vortex-lattice structures emerging as a ground-state solution. Our study focuses on the combined effects of  spin-orbit
coupling and rotation, analysed by using the variational method for single-particle Hamiltonian. We 
observe that the interplay between rotation and Rashba spin-orbit coupling gives 
rise to different effective potentials for the bosons. Specifically, at higher 
rotation frequencies, isotropic spin-orbit coupling leads to an effective toroidal 
potential, while fully anisotropic spin-orbit coupling results in a symmetric double well potential. To obtain these findings, we solve the five coupled Gross-Pitaevskii equations for the spin-2 BEC with spin-orbit coupling under rotation. 
Notably, we find that the antiferromagnetic, cyclic, and ferromagnetic phases 
exhibit similar behavior at higher rotation.
}

\maketitle





\section{\label{sec:level1}Introduction}
In recent decades, advancements in optical traps within cold atom experiments have facilitated investigations into spinor Bose-Einstein 
condensates (BECs). 
The experimental realization of spin-orbit (SO) coupling in spinor BECs has been one of the most important advancements in the last decade 
\cite{lin2011spin}, thus opening up avenues for numerous noval studies on spin-2 BECs \cite{Xu,Kawakami}. 
Spin-2 BECs with SO coupling have also been proposed theoretically \cite{soc-proposals}. In SO-coupled spinor BEC, exotic vortex-lattice configuration has been explored in the presence of rotation \cite{xu}. 
The combined effect
of both SO coupling and rotation frequency facilitates the emergence of a variety of topological configurations
in pseudospin BEC~\cite{radic,xu,zhou,liu1,aftalion,Shi}.
The interplay of isotropic SO coupling and rotation resulting in the emergence
of half-skyrmion excitations in rotating and
rapidly quenched spin-1 BECs using stochastic projected
Gross-Pitaevskii equations have been studied
in \cite{liu3,rotating_spin1-aniso,vortex_lattice-adhikari,necklace_torroidal,weyl_soc}. Rotating spin-1 BEC also supports many interesting 
solutions in the presence of an isotropic \cite{vortex_lattice-adhikari,eff_banger} 
as well as anisotropic, \cite{liu3,rotating_spin1-aniso,eff_banger} SO coupling. Moreover, numerical studies on rotating SO-coupled BEC has been done in 
toroidal traps\cite{necklace_torroidal}, with Weyl SO coupling \cite{weyl_soc}, and SU(3) coupling \cite{su3}.
Recently, by considering the isotropic SO-coupled spin-2 BEC under
rotation topological vortical phase transitions have been studied theoretically \cite{Zhu}. This numerical
study has been done for small to moderate rotation frequency.  However, SO-coupled spin-2  BECs systems 
have not been completely investigated. In this work we fully examine anisotropic as well isotropic SO-coupling with moderate to higher rotation frequencies. 
We investigate a quasi-two-dimensional (q2D) harmonically trapped spin-2 condensate,
featuring an anisotropic SO-coupling term proportional to $S_x \hat{p}_x$, as well as an isotropic SO coupling 
term proportional to {$(S_x \hat{p}_y- S_y \hat{p}_x)$}.
We choose $^{23}$Na  \cite{ciobanu2000phase}, $^{87}$Rb \cite{widera2006precision}, and $^{83}$Rb  
  \cite{ciobanu2000phase} BECs as the prototypical examples of systems with
antiferromagnetic, cyclic, and ferromagnetic spin-exchange interactions. 
We compute the rotational energy per particle 
as a function of rotation frequency, and at higher rotation, antiferromagnetic, cyclic, and ferromagnetic phases exhibit qualitatively similar behavior. 
To analyze the characteristics of the SO-coupled non-interacting Hamiltonian under rotation, we employed a 
variational method. During the investigation of the single-particle Hamiltonian, we interpret the effective 
potentials experienced by bosons. The detailing of calculating effective potentials for SO-coupled spin-2 
BEC under rotation is discussed in Appendix.
The paper is  as follows.
In Section \ref{spd}, we use a variational
method to study the properties of the non-interacting Hamiltonian and interpret the 
effective potential experienced by bosons in the presence of rotation and SO coupling. In Section \ref{MF-model}, we introduce formalism of 
the mean-field model for an SO-coupled spin-2 BEC in the
rotating frame and discuss the Coupled Gross pitiaviskii
Equations (CGPEs).
The Numerical results of the interacting systems are discussed in Section \ref{numerical_result}.
Finally,
in Section \ref{summary}, we conclude the results of our study.
\section{Single particle Analysis}
\label{spd}
In our study, we examine a spin-2 BEC with SO coupling that is confined in the $xy$-plane. The condensate is subjected to a q2D harmonic trap of 
the form $m(\omega_x^2 x^2 + \omega_y^2 y^2+\omega_z^2 z^2)/2$, where $\omega_z$ is significantly larger than $\omega (= \omega_x=\omega_y)$ along the $z$-direction. 
The single particle Hamiltonian for an SO-coupled spin-2 BEC in rotating frame can be described as \cite{radic,kawaguchi2012spinor,goldmann_review}
\begin{equation}
\label{speq}
H = H_{\rm lab} - \Omega L_z
\end{equation}
where $\Omega$ is the angular frequency of rotation, 
$L_z =-i( x\partial/\partial y- y\partial/\partial x) \label{lz}$ is the $z$ component of the angular momentum 
operator, and $H_{\rm lab}$ is single particle laboratory frame Hamiltonian, can be written in the dimensionless form
\begin{equation}
H_{\rm lab}= \left(\frac{\hat{p}_x^2+ \hat{p}_y^2}{2} + \frac{x^2+y^2}{2}\right)\times{\mathds{1}}+ H_{\rm SOC}, 
\end{equation}
where ${\rm{I}}$ represents a $5\times5$ identity matrix, 
and $\hat{p}_d = -i\partial/\partial d$ with $d = x,y$.
The $H_{\rm SOC}$ can be expressed as {\it case I:} $\gamma_x S_x\hat{p}_x$
experimentally achievable equal-strength mixture of Rashba and Dresselhaus couplings \cite{lin2011spin} 
terming as an anisotropic spin-orbit coupling for q2D-system.  Alternatively,  {\it case II:}
$\gamma(S_x\hat{p}_y-S_y\hat{p}_x)$, which represents the Rashba spin-orbit 
coupling \cite{Rashba} between the spin and the linear momentum along the $xy$-plane.
The $\gamma_x$ and $\gamma$ are the SO coupling strength for respective cases.
$S_{d}$ is the irreducible representations of the $d$-component of angular momentum operators for spin-2 matrix with $d=x,y$.
\begin{eqnarray}
(S_x)_{j',j} &=& \frac{1}{2}\left(\sqrt{(6-j'j)}\hbar\delta_{j',j+1}\right.
                 \nonumber\\
             & &+\left.\sqrt{6-j'j)}\hbar\delta_{j'+1,j}\right),\\
(S_y)_{j',j} &=&  \frac{1}{2i}\left(\sqrt{(6-j'j)}\hbar\delta_{j',j+1}\right.
                 \nonumber\\
             & &-\left.\sqrt{(6-j'j)}\hbar\delta_{j'+1,j}\right).
\end{eqnarray} 
with $j'$ and $j$ varies from $-2$ to $2$. To describe the combined effect of  rotation and 
spin-orbit coupling, we analyse the single particle Hamiltonian.
To examine the Eq.~(\ref{speq}) for {\it caseI} we consider the following normalized
variational {\em ansatz}
\begin{equation}\label{c1_ansatz}
 \Psi^{\pm}_{\rm var}=\frac{\exp{\left[-\frac{(x-x_0)^2}{2}-\frac{(y-y_0)^2}{2} +i(k_x x+k_y y)\right]}}{2
 \sqrt{\pi }}
 \times \left(\frac{1}{4}, \mp\frac{1}{2}, \sqrt{\frac{3}{8}},\mp\frac{1}{2}, \frac{1}{4}\right)^T ,
\end{equation}
where $x_0,y_0,k_x$ and $k_y$ are the variational parameters. The variational energies are
\begin{subequations}
\begin{align}
E^{\pm}_{\rm var}(x_0,y_0,k_x,k_y) 
&= \int dx dy{\Psi^{\pm}_{\rm var}}^* {\color{blue}H} \Psi^{\pm}_{\rm var},\nonumber\\
 &=\frac{1}{2} \big[k_x (\mp4 \gamma_x+k_x+2 \Omega y_0)+k_y^2\nonumber\\&-2 k_y x_0\Omega +x_0^2+y_0^2+2\big],
\end{align}
\end{subequations}
where ${\Psi_{\rm var}^{\pm}}^* $ is the conjugate transpose of $\Psi_{\rm var}^{\pm}$.
The variational parameters can be fixed by minimizing $E^{\pm}_{\rm var}$ with respect to 
$(x_0,y_0,k_x,k_y)$. The location(s) of minima thus obtained are 
\begin{equation}
x_0 = 0,~y_0 = \mp\frac{2\gamma_x \Omega}{1-\Omega^2},
~ k_x = \pm \frac{2\gamma_x}{1-\Omega^2},~ k_y = 0,\label{var_parameters_aniso}     
\end{equation}
\begin{equation}
  E^{\pm}_{\rm min} =\frac{2 \gamma_x^2+\Omega ^2-1}{\Omega^2-1}\label{emin_a},
\end{equation}
for $\Psi_{\rm var}^{\pm}$.
The most generic variational solution for these parameters' sets is 
$c_{+}\Psi_{\rm var}^+ + c_{-}\Psi_{\rm var}^-$, where $c_{\pm}$ are coefficients of superposition with 
${\vert c_{+}\vert}^2+\vert{c_{-}\vert}^2=1$. The resultant density profile reflects 
the effective two-well potentials with two minima at 
$(x_0 = 0, y_0 =  \mp{2\gamma_x \Omega}/{1-\Omega^2})$. 
The validity of the variational method has been checked by considering the set of parameter $(\gamma_x, \Omega )$ is $(1,0.5)$, $(1,0.7)$, and $(1,0.9)$. 
For these parameters set, the variational solutions $\Psi_{\rm var}^{\pm}$ are degenerate, and the peak of total 
variational densities lies at ${\pm 1.33,~\pm 2.74,~\pm 9.47}$, 
respectively. The densities profiles obtained by the variational method for non-interacting systems are in 
excellent agreement with the numerical results (not shown here) and so as the corresponding energies.
The rotating spin-2 BEC in the presence of ansitropic SO-coupling coupling is exactly solvable \cite{eff_banger}.
The ansatz presented here is an exact analytic solution to a one-dimensional SO coupling Hamiltonian under rotation.
 
\begin{figure*}[ht]
\centering
\includegraphics[width=0.98\textwidth]{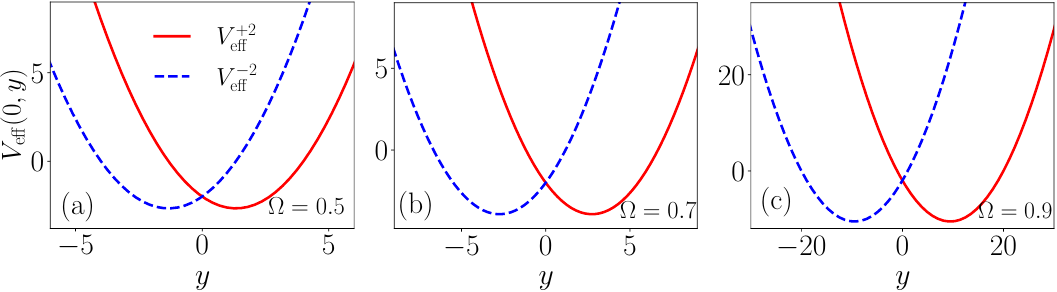}
\caption{(Color online) 
(a)-(c) are the curves of effective potentials
corresponding to Eq.~(\ref{Veff}) for 
$\Omega_{\rm rot}=0.5,0.7$ and 0.9, respectively.
For these curves, SO-coupling strength is
$\gamma_x=1$.}
\label{effpot}
\end{figure*}
The effective potential incurred by the boson can be evaluated by calculating the vector
and scalar potentials as discussed in ref. \cite{radic,eff_banger}.
With the definition given in \cite{radic},
the effective potential for rotating SO-coupled spin-2 system calculated in Appendix.  The effective potential curves for the system are 
\begin{align}\label{veffj2}
V_{\rm eff}^{j}(x,y) &= \frac{1}{2}\left[ (1-\Omega^2)(x^2+y^2)-j^2\gamma_x^2-2 j \gamma_x\Omega y\right].
\end{align}
 where $j =\pm 2,\pm1, 0$.
The minima of effective potential occurred in the curves corresponding to 
$j=\pm2$ depending on the region of $y>0$ or $y<0$, respectively illustrated in 
Fig.~\ref{effpot}. Eq.~(\ref{veffj2}) exhibiting that $V_{\rm eff}^{+2}(x,y)$ and $V_{\rm eff}^{-2}
(x,y)$ are overlap at $(x=0,y=0)$ as shown in Fig.~\ref{effpot}. In the region $y<0$,  $V_{\rm eff}^{-2}$ corresponding to 
$j=-2$ curve is lower than other for $j=0,\pm1,+2$ (curves for $j=0,\pm1$ are not shown here) and in  $y>0$,  $V_{\rm eff}^{+2}$ which is corresponding to 
$j=+2$ is lower than the other curves for $j=0,\pm1,-2$. 
Fig.~\ref{effpot}(a)-(c) show the effective potential curves for rotation frequencies $\Omega=0.5, 0.7$ and 
$0.9$, respectively.
The potentials incurred by  boson are effectively equivalent to symmetric double-well potentials with minima
exhibit at $(x=0,y=\mp1.33)$,  $(x=0,y=\mp2.74)$, and $(x=0,y=\mp9.47)$ for
$\Omega=0.5, 0.7$ and $0.9$, respectively.

In order to investigate the system with isotropic SO coupling, we consider the following variational 
{\em ansatz}
\begin{equation}
\begin{split}
\Phi_{\rm var}=&\frac{\exp{ \left(\frac{r^2}{2 \sigma ^2}\right)}}{\sqrt{\pi \sigma ^{2 n+4} \Gamma 
(n+2)}}\times(A_1 r^{\vert n \vert} e^{i n\phi }, A_2 r^{\vert n+1\vert} e^{i (n+1)\phi }, 
 A_3 r^{\vert n+2 \vert}e^{i(n+2)\phi},\\&A_4 r^{\vert n+3\vert} e^{i (n+3)\phi }, 
A_5 r^{\vert n+4\vert}e^{i(n+4)\phi})^T,\label{c2-ansatz}
\end{split}
\end{equation}
where, the variational amplitudes are denoted as $A_1, A_2, A_3, A_4, A_5$, while $\sigma$ represents the 
variational width of the ansatz, and $n$ is a variational integer.
The normalization condition imposes the constraint
\begin{multline}
(n+2) \sigma ^2 \left((n+3) \sigma ^2 \left(A_5^2 (n+4) \sigma ^2+A_4^2\right)+A_3^2\right)+A_2^2+\frac{A_1^2}{(n+1) \sigma ^2}=1,\label{norm_con}
\end{multline}
on the variational parameters $A_1,A_2,A_3,A_4,A_5,n$ and $\sigma$. The variational energy is
\begin{multline}
E_{\rm var}(A_1,A_2,A_3,A_4,A_5,n,\sigma)=[4 A_1 A_2 \gamma  (n+1) \sigma ^2+(n+1) \sigma ^2 (2 \sqrt{6} A_2 A_3 \gamma  
(n+2) \sigma ^2\\+(n+2) \sigma ^2 (2 \sqrt{6} A_3 A_4 \gamma  (n+3) \sigma ^2+(n+3) \sigma ^2 (4 A_4 A_5 \gamma  (n+4) \sigma ^2+A_5^2 (n+4) \sigma ^2 ((n+5) \\
(\sigma ^4+1)-2 (n+4) \sigma ^2 \Omega )+A_4^2 ((n+4) (\sigma ^4+1)-2 (n+3) \sigma ^2 \Omega ))+A_3^2((n+3) (\sigma ^4+1)-\\2 (n+2) \sigma ^2 \Omega 
))+A_2^2((n+2)(\sigma ^4+1)-2 (n+1) \sigma ^2 \Omega))+A_1^2 ((n+1)(\sigma 
^4+1)\\-2 n \sigma ^2 \Omega )]/(2(n+1)\sigma^4) \label{Evar_iso}
\end{multline}
The variational energy can be minimized with all variational parameters subjected to the constraint specified in Eq (\ref{norm_con}), to decide the variational parameters. To check the validity of 
the variational method in this case, we choose
$(\gamma = 1,{\color{blue}\Omega }= 0.9)$,
the variational parameters while minimization of (\ref{Evar_iso}) are 
$(A_1,A_2,A_3,A_4,A_5,n,\sigma) = (-2.3728,0.5012,-0.0645,0.0055,~-0.0002,98,0.9491)$.
The comparison of variational density $\vert \Phi_{\rm var}(r)\vert^2$, and exact numerically evaluated single
particle density profiles $\vert\Psi(r)\vert^2$,  agree with each other.
The peak of total variational density lies 
along a circle of radius $9.47$ and is related
to the effective toroidal potential incurred by the boson. 
For $(\gamma = 1, \Omega = 0.7)$ the variational parameters while minimization are variational energy 
$(A_1,A_2,A_3,A_4,A_5,n,\sigma) =(0.7029,-0.5115,0.2173,-0.0589,0.0094,9,0.8489 )$
for parameters' sets, respectively. The comparison of 
$\vert \Phi_{\rm var}(r)\vert^2$, and $\vert\Psi(r)\vert^2$, agree with each other and the peak of total variational density lies 
along a circle of radius $2.70$. 
\section{Mean-field interacting Model}
\label{MF-model}
In q2D trapping potential, a rotating SO-coupled spin-2 BEC under mean-field approximation can be described by a set of five CGPEs \cite{kawaguchi2012spinor} 

\begin{subequations}
\begin{eqnarray}
i\frac{\partial \phi_{\pm 2}}{\partial t} &=& \mathcal{H} \phi_{\pm 2} + 
c_0 {\rho} \phi_{\pm 2} + c_1 \{F_{\mp}\phi_{\pm 1} 
\pm 2 F_{z} \phi_{\pm 
2}\}+c_2 \frac{\Theta \phi_{\mp 2}^*}{\sqrt{5}}+ \Gamma_{\pm 2}, 
\label{cgpet2d-1}\\
i\frac{\partial \phi_{\pm 1}}{\partial t} &=& \mathcal{H} \phi_{\pm 1}
+ c_0 {\rho}\phi_{\pm 1} + c_1 \left(\sqrt{\frac{3}{2}} F_{\mp}
\phi_{0} +F_{\pm} \phi_{\pm 2}\pm F_{z} \phi_{\pm 1} \right) \nonumber\\
&&
- c_2 \frac{\Theta \phi_{\mp 1}^*}{\sqrt{5}} + \Gamma_{\pm 1}, 
\label{cgpet2d-2}\\
i\frac{\partial \phi_0}{\partial t} &=& \mathcal{H} \phi_0 
+ c_0 {\rho}
\phi_0 + c_1 {\sqrt{\frac{3}{2}}}\{F_{-} \phi_{-1} +  
F_{+} \phi_1\}+ c_2 \frac{\Theta \phi_{0}^*}{\sqrt{5}}+\Gamma_0, \label{cgpet2d-3}
\end{eqnarray}
\end{subequations}
where, $\Phi=(\phi_2,~\phi_1,~\phi_0,~\phi_{-1},~\phi_{-2})^T$ is a five component  order parameter, 
\begin{eqnarray*}
\mathcal{H}&=&-\frac{\mathbf{\nabla^2}}{2}+V -\Omega L_z,\quad \Theta = 
\frac{2\phi_2 \phi_{-2} - 2\phi_1\phi_{-1}+ \phi_0^2}{\sqrt{5}},
\quad F_z = \sum_{j=-2}^2 j\mid \phi_j\mid^2\\
F_- &=& F_+^* = 2\phi_{-2}^* \phi_{-1} + \sqrt{6}\phi_{-1}^* 
\phi_0 + \sqrt{6} \phi_0^* \phi_1 + 2 \phi_2 \phi_1^*, 
\end{eqnarray*}
and $\rho = \sum_{j=-2}^2 \vert\phi_j\vert^2$ is the total density. The Laplacian, trapping
potential $V$, interaction parameters $(c_0,c_1,c_2)$,
are defined as
\begin{subequations}
\begin{eqnarray}
\mathbf{\nabla^2} &=& \left(\frac{\partial}{\partial x^2} + \frac{\partial}{\partial y^2}
\right),~V = {\frac{ x^2 
+ y^2}{2}},~
c_0 = \sqrt{2\pi\alpha_z}\frac{2\pi N(4 a_2 +3 a_4)}{7 a_{\rm osc}},\\
c_1& =& \sqrt{2\pi\alpha_z}\frac{2\pi N( a_4 - a_2)}{7 a_{\rm osc}},\quad
c_2 = \sqrt{2\pi\alpha_z}\frac{2\pi N( 7 a_0 - 10 a_2 +3a_4 )}{7 a_{\rm osc}},
\end{eqnarray}
\end{subequations}
and the $\Gamma$'s for {\it case I} are
\begin{subequations}
\begin{align}
\Gamma_{\pm 2}&=-i\gamma_x\frac{\partial\phi_{\pm 1}}
{\partial x},\quad
\Gamma_{\pm 1}=-i \left(  \gamma_x\frac{\partial\phi_{\pm 2}}{\partial x} 
+\sqrt{\frac{3}{2}}
\gamma_x\frac{\partial\phi_0}{\partial x} \right) ,\nonumber\\
\Gamma_{0}&=-i\left(\sqrt{\frac{3}{2}}
\gamma_x\frac{\partial\phi_1}{\partial x} +\sqrt{\frac{3}{2}}\gamma_x
\frac{\partial\phi_{-1}}{\partial x} \right), \nonumber
\end{align}
\end{subequations} 
for {\it case II} are
\begin{subequations}
\begin{align}
\Gamma_{\pm 2}&=-i\gamma\left(\frac{\partial\phi_{\pm 1}}
{\partial y}\pm i\frac{\partial\phi_{\pm 1}}{\partial x}\right),\nonumber\\
\Gamma_{\pm 1}&=-i \gamma\left(  \frac{\partial\phi_{\pm 2}}{\partial y} 
+\sqrt{\frac{3}{2}}
\frac{\partial\phi_0}{\partial y} 
\mp i\frac{\partial\phi_{\pm 2}}
{\partial x} \pm i\sqrt{\frac{3}{2}}\frac{\partial\phi_0}{\partial x} 
\right),\nonumber\\
\Gamma_{0}&=-i\gamma\left(\sqrt{\frac{3}{2}}
\frac{\partial\phi_1}{\partial y} +\sqrt{\frac{3}{2}}
\frac{\partial\phi_{-1}}{\partial y} 
- i{\sqrt{\frac{3}{2}}}\frac{\partial
\phi_1}{\partial x} +i\sqrt{\frac{3}{2}}\frac{\partial\phi_{-1}}
{\partial x}\right). \nonumber
\end{align}
\end{subequations}
where $a_0, a_2, a_4$ represent the $s$-wave scattering lengths in the allowed scattering channels, and $\alpha_x,\alpha_y$ and $\alpha_z$ are the anisotropy parameters of trapping frequency.
All these quantities  are dimensionless, which have been calculated by expressing
lengths ($a_0,a_2,a_4,x,y$)
in units of $a_{\rm osc} \equiv\sqrt{\hbar m/ \omega}$, energy, density and time in the unit  
of $\hbar \omega$, $a_{\rm osc}^{-2}$ , and  $\omega^{-1}$, respectively.
The dimensionless form of the mean-field model for the
SO-coupled spin-2 has two conserved quantities: one is the normalization condition 
 $ \int \rho(x, y)dxdy=1$. And the other is energy per particle defined as
\begin{eqnarray}\label{energypp}
E &=& \int dx dy
\left[\sum_{j=-2}^{+2}\phi_j^{*}\mathcal{H}\phi_j
+ \frac{c_0}{2}{\rho}^2  + \frac{c_1}{2}\vert {\bf F}\vert^2 + 
\frac{c_2}{2}\vert \Theta \vert ^2 +\sum_{j=-2}^{+2}\phi_{j}^*\Gamma_{j} \right],
\end{eqnarray}
for an SO-coupled BEC.
Under the influence of SO coupling longitudinal magnetization 
\begin{equation}
    {\cal M} = \int \left( 2\vert\phi_{+2}\vert^2 +\vert \phi_{+1}\vert ^2-\vert \phi_{-1}\vert ^2-2 \vert \phi_{-2}\vert^2\right) dx dy,
\end{equation}
is not conserved, although it remain
conserved for $\gamma_x=0$ and $\gamma=0$.

\section{Numerical Results}
\label{numerical_result}
\begin{figure*}[ht]
\centering
\includegraphics[width=0.7\textwidth]{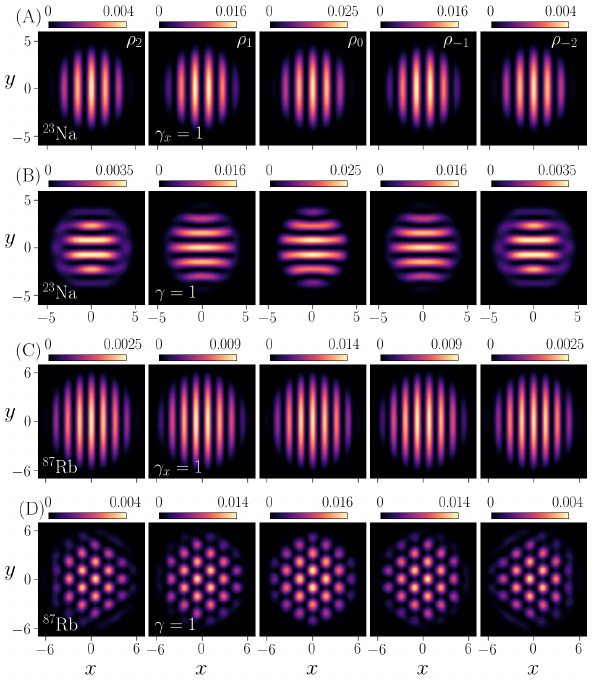}
\caption{(Color online) 
(A) and (B) shown the ground state component densities for an anisotropic SO coupling  and isotropic  SO-coupling  $^{23}$Na spin-2 BEC, respectively. The interaction parameters for both cases are $c_0=340.45,c_1=16.90, c_2=-18.25$, and SO coupling strength for (A) $\gamma_x=1$ and (B) = $\gamma=1$.
Similarly, (C) and (D) represent the same for $^{87}$Rb spin-2 BEC for interaction parameters $c_0=1164.80,c_1=13.88, c_2=0.43$.
}
\label{intial_sol}
\end{figure*}
We take a typical example of antiferromagnetic system trapped in a q2D potential, consisting of $50,000$ atoms of an SO-coupled $^{23}$Na spin-2 BECs 
with $\omega = 2\pi\times 10$ Hz, $\omega_z = 2\pi\times 100$ Hz, and
so $\alpha_x = \alpha_y = 1$ and $ \alpha_z = 10$ where $\alpha_{\nu}=\omega_{\nu}/\omega_x$.
The three scattering lengths considered for this system are 
$a_0 =34.9a_B, a_2 = 45.8a_B, a_4 = 64.5a_B$ \cite{ciobanu2000phase}
and corresponding triplet of dimensionless interaction 
strengths are
$(c_0,~c_1,~c_2) =(340.45,~ 16.90,~-18.25)$. The oscillator length for this system is $4.69\mu$m.
For the cyclic phase, we examine a system composed of $50,000$ atoms of $^{87}$Rb spin-2 BECs confined in a q2D trapping potential with the same trapping frequencies considered for antiferromagnetic phase. The chosen triplet of scattering lengths for this system consists of $a_0 = 87.93a_B$, $a_2 = 91.28a_B$, and $a_4 = 99.18a_B$ \cite{widera2006precision}, resulting in the corresponding dimensionless interaction strengths of $(c_0, c_1, c_2) = (1164.80, 13.88, 0.43)$. The oscillator length in this system is $2.41\mu$m.

We investigate the stationary state solutions of these systems in the presence SO coupling.
To solve the set of CGPEs in (\ref{cgpet2d-1})-(\ref{cgpet2d-3})), we employ the time-splitting Fourier spectral method \cite{spin1-soc,banger2021semiimplicit,ravisankar-cpc,spinf-soc}.
To obtain the numerical solutions, a spatial step size of $0.1$ and a temporal step size of $0.001$ to be used. In the absence of a rotation frequency 
($\Omega= 0$), the equations are evolved in imaginary time propagation to obtain the ground 
state of the system. Multiple numerical simulations are performed with different initial guesses, and 
the state with the lowest energy is considered as the ground state solution. When $\Omega = 0$, the spin-2 BEC can exhibit various ground state 
solutions depending on the interaction parameters and the strength of the spin-orbit coupling.
The most general solution is axisymmetric, characterized by $(-2,-1,0,+1,+2)$ phase singularities at densities corresponding to 
$j=+2,+1,0,-1,-2$ components \cite{spin-2-PK}. Additionally, other patterns such as stripes \cite{spin-2-PK}, square lattices \cite{spin-2-PK}, 
or triangular lattices \cite{spin-2-PK} can arise through the superposition of counter-propagating 
plane waves, four plane waves with propagation 
vectors at a right angle to each other, or three plane waves with propagation vectors at an angle of $2\pi/3$ to each other, respectively.
When performing simulation in a rotating frame, previously obtained ground state solutions are 
considered as an initial guess.
The dynamical stability of each solution has been studied through real-time propagation, with the 
converged solution obtained from the imaginary time propagation serving as the initial guess for these 
dynamical stability studies.

For {\it case I}
In the absence of rotation frequency $(\Omega=0)$,  
numerically obtained ground state solutions for q2D configuration with the aforementioned 
coupling for $\gamma_x=1$  having 
a vertical stripe pattern in the component densities shown in Fig. \ref{intial_sol}A and  \ref{intial_sol}C
for antiferromagnetic and cyclic phase, respectively. {The vertical stripe are emerge due to the effect of SO-coupling $\propto S_x{\hat p}_x$.}
Fig. \ref{intial_sol}B and  \ref{intial_sol}D, display the ground state solutions for {\it case II}, representing horizontal stripe patterns in the 
antiferromagnetic and triangular-lattice 
pattern in cyclic phase, respectively, for the aforementioned 
coupling strength $\gamma=1$. { The nature of the solution would depend on the combined effect of type of interaction, type of SO coupling and strength of SO coupling.} 
\subsection*{Antiferromagnetic phase}
\begin{figure*}[ht]
\centering
\includegraphics[width=0.8\textwidth]{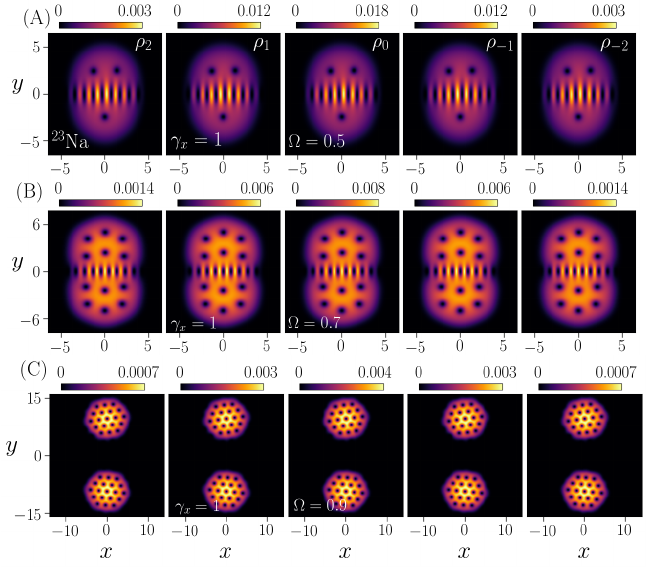}
\caption{(Color online) 
(A) shows the ground state component densities for an anisotropic SO-coupled
$^{23}$Na spin-2 BEC with $c_0=340.45,~c_1=16.90,~c_2=-18.25$, 
$\gamma_x=1$  with $\Omega =0.5$.
(B) and (C) represent the same for the rotation frequencies $\Omega=0.7$ and 0.9, respectively.
}
\label{fig:figpa}
\end{figure*}
\begin{figure*}[ht]
\centering
\includegraphics[width=0.8\textwidth]{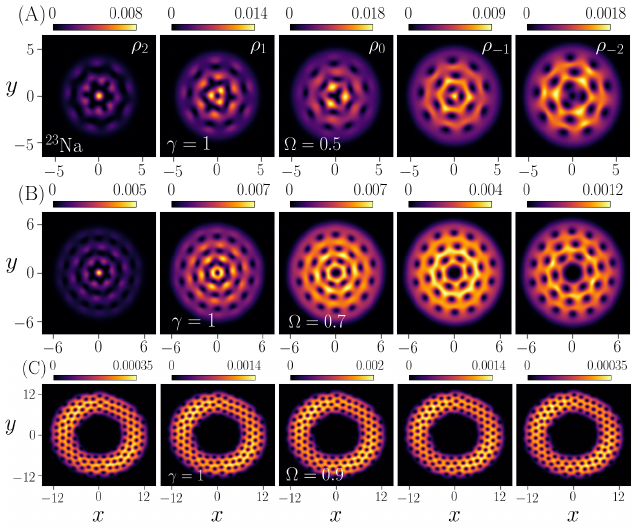}
\caption{(Color online) 
(A) present the component densities of the ground state for an isotropic SO-coupled
$^{23}$Na spin-2 BEC with $c_0=340.45,c_1=16.90, c_2=-18.25$, $\gamma=1$  with $\Omega =0.5$.
 (B) and (C) showing the same for rotation frequencies $\Omega=0.5,~0.7$, and 0.9, respectively.
}
\label{fig:figpi}
\end{figure*}
{\it Case I:} For the simulation in the rotating frame, we consider different rotation frequencies, $\Omega=0.5,0.7,$ and $0.9$ times the trapping frequency. We select these rotation frequencies to explore all possible vortex lattice configurations that can arise.
The anisotropy in the SO coupling leads to a distinction in the angular momentum of the components along the $x-$ and $y$-directions, thereby impacting the rotational symmetry.
Fig. \ref{fig:figpa}(A) illustrates the component densities with a rotation frequency of $\Omega=0.5$, where a significant portion of the vortices align themselves in a central chain along the x-direction.
When considering higher rotation frequencies, such as $\Omega=0.7$, a larger number of vortices emerge in the system. Some vortices appear on both sides of the central chain, as depicted in Fig. \ref{fig:figpa}(B).  The number of vortices in the
off-vortices-chain region increases with the rotation frequency. At very large rotation frequency $\Omega=0.9$
all the vortices arrange themselves on both sides of the central chain, as illustrated in Fig. \ref{fig:figpa}(C). Here, we observed that with strong anisotropic SO coupling and large rotation, some local minima about the trap become discernible, and triangular vortex lattices would preferentially form in the regions outside the vortex chain.

{\it Case II:} In the rotating frame, we have observed the evolution of the solution already obtained with $(\Omega=0)$ under various rotation frequencies, 0.5, 0.7, and 0.9. 
Fig. \ref{fig:figpi}(A) illustrates the solution rotated with a frequency of $\Omega=0.5$.
With this relatively low rotation frequency, a small number of 
vortices have emerged, and all of them tend to arrange themselves in a regular pattern.
Fig. \ref{fig:figpi}(B) depicts the vortex lattice structure obtained with a higher rotation frequency of $\Omega=0.7$. In this
case, the central region of the structure accommodates singularities with charges $(0,+1,+2,+3,+4)$ in the $(j=-2,-1,0,+1,+2)$ components, respectively. The charges of 
these singularities have been determined from the phase of the density profiles (not shown here).
As the rotation frequency increases, the number of vortices also increases. 
Subsequently, we further increased the rotation frequency to $\Omega=0.9$. At this 
frequency, we observed the formation of a ring-type structure with a giant vortex
located near the center of the trap. And triangular vortex lattices appear in the annular region surrounding the central vortex depicted in Fig. \ref{fig:figpi}(C).
The number of vortices within a circle of radius $9.47$ is consistent with the number obtained from variational
analysis. The variational analysis for this case is showing a excellent agreement particularly at higher 
rotation frequencies.
\subsection*{Cyclic phase}
\begin{figure*}[ht]
\centering
\includegraphics[width=0.7\textwidth]{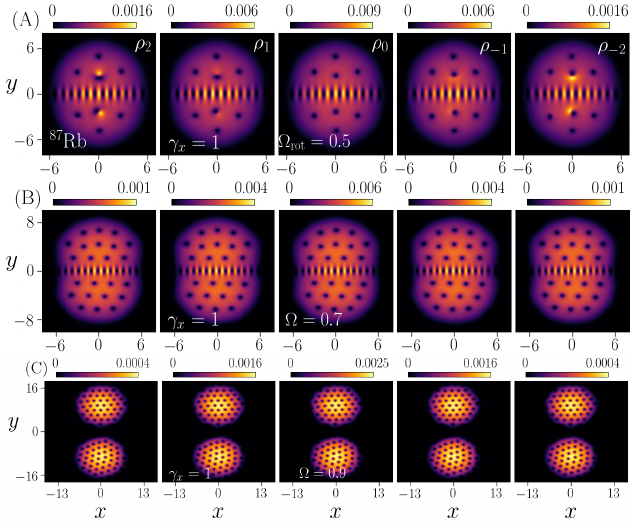}
\caption{(Color online) 
(A) shows the ground state density profiles of the individual components
of the SO-coupled $^{87}$Rb spin-2 BEC with $c_0=1164.80,~c_1=13.88,~c_2=0.43$, anisotropic SO coupling $\gamma_x=1$, $\Omega =0.5$. Similarly,
(B) and (C) present the component densities for $\Omega =0.7$ and 0.9, respectively.
}
\label{fig:figca}
\end{figure*}
\begin{figure*}[ht]
\centering
\includegraphics[width=0.7\textwidth]{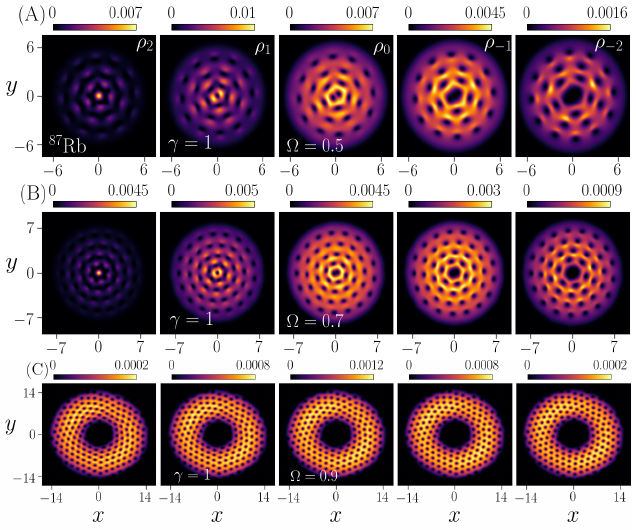}
\caption{(Color online) 
(A) shows the ground state density profiles of the individual components
of the SO-coupled $^{87}$Rb spin-2 BEC with $c_0=1164.80,~c_1=13.88,~c_2=0.43$, isotropic SO coupling $\gamma=1$, $\Omega =0.5$. Similarly,
(B) and (C) present the component densities for $\Omega =0.7$ and 0.9, respectively.
}
\label{fig:figci}
\end{figure*}
\begin{figure*}[ht]
\centering
\includegraphics[width=0.5\textwidth]{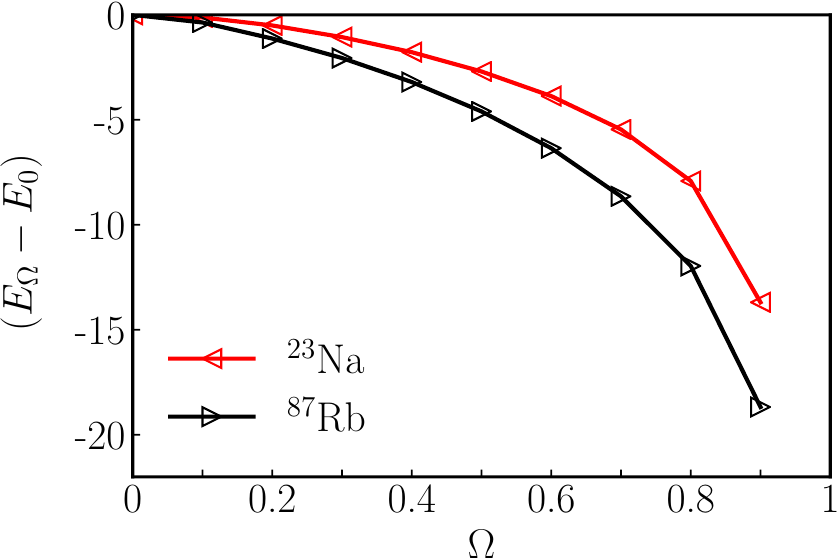}
\caption{(Color online) The plot shows the variation of rotational energy $(E_{\Omega}-E_0)$ with the $\Omega$ for both antiferromagnetic $^{23}$Na and cyclic $^{87}$Rb systems. 
}
\label{energydiff}
\end{figure*}

{\it Case I:} The solution reported in Fig. \ref{intial_sol}(C) evolved in a rotating frame with rotation frequencies of $\Omega=0.5$, $0.7$, and $0.9$. The corresponding results are depicted in Fig. \ref{fig:figca}(A), (B), and (C), respectively.
At a small rotation frequency of $\Omega=0.5$, the majority of vortices align 
themselves along a horizontal chain, while the remaining vortices position themselves
either above or below this vortex chain. This configuration is illustrated in 
Fig. \ref{fig:figca}(A). As the rotation frequency increases to $\Omega=0.7$, 
more vortices emerge in the condensate, as depicted in Fig. \ref{fig:figca}(B).
Finally, at a very high rotation frequency of $\Omega=0.9$, a large number of 
vortices are observed, and they arrange themselves in a triangular vortex lattice 
pattern, as shown in Fig.~\ref{fig:figca}(C).

{\it Case II:}  Fig. \ref{fig:figci}(A) illustrate the
vortex lattice configuration at $\Omega=0.5$.
In Fig. \ref{fig:figci}(B), vortex lattice configuration shows corresponding to $\Omega=0.7$, where 
the central part of the structure hosting $(0, +1,+2,+3,+4)$ 
singularities in $(j=-2,-1,0,+1,+2)$ components. 
At $\Omega=0.9$, a vortex lattice structure having a giant vortex at the center and the triangular vortex lattice pattern appear in the annulus as shown in Fig. \ref{fig:figci}(C).

The rotational energy  of a scalar BEC  in the
rotating frame $(E_{\Omega}-E_0)$ is the energy of rigid-body rotation $-I\Omega^2/2$
where $I$ is the moment of inertia of the condensate \cite{Fetter}.
This energy
is proportional to the square of the angular frequency \cite{Fetter,vortex_lattice-adhikari}, where $E_{\Omega}$ is given 
by Eq. (\ref{energypp}). In the case of spin-2 spinor antiferromagnetic and cyclic BECs, also hold a similar relation.
We illustrate in Fig.\ref{energydiff} the rotational energy for both antiferromagnetic and cyclic spin-2 BEC for isotropic SO coupling. Here, we plot
the energy of the ground state versus $\Omega$.
The energies of  antiferromagnetic and cyclic systems lie in two separate
distinct lines exhibit similar qualitative behavior.
As the rotation frequency increases, the energy decreases due to the negative contribution of the rotational energy term $-\Omega L_z$ in the energy expression Eq. (\ref{energypp}). 
This behavior is particularly evident in Figure \ref{energydiff}, where the rotational energy is behaving $\propto$ $-\Omega^2$ as the rotation frequency increases specially at higher $\Omega$.

{\it Ferromagnetic phase:}
Here we have considered $50,000$ atoms of $^{83}$Rb
in an SO-coupled spin-2 BEC with or without rotation frequency. The value of three scattering length considered for this system are $a_0 =83.0a_B, a_2 = 82.0a_B, a_4 = 81.0a_B$ 
\cite{ciobanu2000phase} and corresponding interaction parameters are 
$(c_0,c_1,c_2) = (980.33,-1.71,6.86)$.
The numerical results obtained for $^{83}$Rb under rotation demonstrate 
qualitative similarities to the cyclic $^{87}$Rb BEC. So as we have 
opted not to include the results obtained for the ferromagnetic case
in this work.  In the actual parameters magnitude of $c_1$ and $c_2$ 
is very small comparing to $c_0$, so at moderate to high rotation frequencies the
effect of interaction is neglible compare to $\Omega$. 
By adjusting the parameters $c_1$ (more attractive) or $c_2$ (more repulsive) or both, such that the system's interaction becomes 
comparable to the rotation frequency, we can effectively study the impact of interaction terms. This will allow us to distinguish 
between the ferromagnetic and cyclic phases.
\cite{eff_banger}.
\section{summary}
\label{summary}
We have demonstrated the vortex lattice structures in q2D in 
SO-coupled spin-2 BEC under rotation by employing the variational method for single particle
and numerical solution of the mean-field model. 
In this work, we considered isotropic as well as fully anisotropic SO-coupling strengths with moderate
and higher rotation frequencies. In this work,
antiferromagnetic $^{23}$Na , cyclic phase  $^{87}$Rb and ferromagnetic $^{83}$Rb have been discussed in detail.
We observed that at fast rotational frequencies, all the 
phases have qualitatively similar behavior
which highlights the
negligible role of the spin-exchange interactions.
In the presence of anisotropic SO coupling $\gamma_x S_x \hat{p_x}$ bosons experienced symmetric double-well potential,
and the central vortices chain appeared along the $x$-direction in the numerical solutions.
Depending on the rotation frequency, 
the majority of vortices are arranged on both sides of the central chain of vortices.
We have reported effective potential curves
explored by using single particle Hamiltonian.
With isotropic SO coupling {$\gamma(S_x \hat{p_y}-S_y \hat{p_x})$} in the presence
of rotation frequency
bosons experienced toroidal shape trapping potential, and the radius of the toroidal trap increased with the rotation frequency.
So at higher rotation frequency in an isotropic SO-coupled system giant vortex-type solution 
observed as a ground state solution. At large rotation, vortices favor triangular vortex lattice patterns in the annual region.
All the numerical results reported in this work have been explained by using variational 
analysis for single particle Hamiltonian.

\section{Appendix}
The single particle Hamiltonian for case I is
\begin{equation}
H_{\rm rot}= \left(\frac{\hat{p}_x^2+ \hat{p}_y^2}{2} + \frac{x^2+y^2}{2}-\Omega L_z\right)\times{\mathds{1}}+\gamma_x S_x p_x, 
\end{equation}
Under a unitary transformation the $H_{\rm rot}$ can be diagonalised, which has the eigenvalues 
\begin{subequations}\label{en_sph}
\begin{eqnarray}
E_{\pm2}(k_x,k_y) &=&\frac{1}{2} \left[k_x^2+k_y^2 \pm 4\gamma_xk_x\right]
-\Omega(xk_y-yk_x),\\
E_{\pm1}(k_x,k_y) &=&\frac{1}{2} \left[k_x^2+k_y^2 \pm 2\gamma_xk_x\right]-\Omega(xk_y-yk_x),\\
E_0(k_x,k_y) &=& \frac{1}{2}(k_x^2+k_y^2)-\Omega(xk_y-yk_x),
\end{eqnarray}
\end{subequations}
The spectrum in Eqs.~(\ref{en_sph}a)-(\ref{en_sph}b) around a minima can be 
described  by a parabola of form $(k_x-k_{x\rm min})^2+(k_y-k_{y\rm min})^2+E_{\rm min}$,
which describes the particle moving in an effective gauge field $({\bf A},\bar{\Phi})=(\{k_{x\rm min},k_{y\rm min},0\},E_{\rm min})$, where ${\bf A}$ and $\bar{\Phi}$ are vector and scalar potentials, respectively \cite{radic}.
In the presence of rotation term $k_{x\rm min}, k_{y\rm min}$ and $E_{\rm min}$ become spatially dependent \cite{radic}. Rewriting Eqs.~(\ref{en_sph}a)-(\ref{en_sph}b) as
\begin{subequations}
\begin{eqnarray}
E_{\pm2}(k_x,k_y) &=&\frac{1}{2} \left[(k_x\pm(2\gamma_x \pm y\Omega))^2
               +(k_y-x\Omega)^2\right]\nonumber\\ &&
               +E_{\pm2}^{\rm min}(x,y),\\
E_{\pm1}(k_x,k_y) &=&\frac{1}{2} \left[(k_x\pm(\gamma_x \pm y\Omega))^2
               +(k_y-x\Omega)^2\right]\nonumber\\ &&
               +E_{\pm1}^{\rm min}(x,y),\\
E_0(k_x,k_y) &=& \frac{1}{2}((k_x+y\Omega)^2+(k_y-x\Omega)^2)\nonumber\\
             &&+E_0^{\rm min}(x,y),
\end{eqnarray}
\end{subequations}
where ${\bf A}_{\pm2} = \mp (2\gamma_x\pm y\Omega,x\Omega),\quad{\bf A}_{\pm1} = \mp (\gamma_x\pm y\Omega,x\Omega),\quad {\bf A}_0 = (-y\Omega,x\Omega)$, and $\bar{\Phi}_j = E_j^{\rm min}$ with ($j = 0,\pm 1,\pm 2$).
The effective potentials, which are the sums of trapping and scalar potentials \cite{radic}, can now be written as
\begin{subequations} \label{Veff}
\begin{align}
V_{\rm eff}^{\pm2}(x,y) &= \frac{1}{2}\left[ (1-\Omega^2)(x^2+y^2)-4\gamma_x^2 \mp 4y \gamma_x\Omega\right],\\
V_{\rm eff}^{\pm1}(x,y) &= \frac{1}{2}\left[ (1-\Omega^2)(x^2+y^2)-\gamma_x^2\mp 2y \gamma_x\Omega\right],\\
V_{\rm eff}^0(x,y) &= \frac{1}{2}\left[ (1-\Omega^2)(x^2+y^2)\right].
\end{align}
\end{subequations}
From Eqs. (\ref{Veff})a - (\ref{Veff})c, the effective potential curves correspond to 
$V_{\rm eff}^{\pm2}$ are the lowest lying depending on the region.

\section {Acknowledgement}
I thank Sandeep Gautam for the fruitful discussions and comments on the manuscript.

\end{document}